\documentstyle[aps,prl]{revtex}
\begin{document}
\draft
\twocolumn[
\title{Cluster dynamics in systems with constant mean field coupling}
\author{ A. Parravano and M. G. Cosenza}
\address{Centro de Astrof\'{\i}sica Te\'orica,
Universidad de Los Andes,  Apartado Postal 26 La~Hechicera, 
M\'erida~5251, Venezuela.}
\date{Submitted to Int. J. Bifurcations and Chaos, 1998}
\maketitle
\begin{abstract}
\widetext
A procedure to predict the occurrence of periodic clusters in a system 
of globally coupled maps displaying a constant mean field is presented. 
The method employs the analogy between a system of globally coupled maps and 
a single map driven by a constant force.  
By obtaining the asymptotic orbits of the driven map, an associated
coupling function can be constructed. This function allows to
establish a direct connection between both systems.
Some applications are shown.
\narrowtext
\end{abstract}
]

The phenomenon of emergence of nontrivial collective behaviors
in dynamical systems of
interacting chaotic elements has been a focus of recent attention
[Kaneko, 1990; Chat\'e \& Manneville, 1992(a), 1992(b); 
Chat\'e {\it et al.}, 1996; Chawanya \& Morita, 1998].
An important class of systems 
which can exhibit ordered collective effects are globally coupled nonlinear
oscillators. Such systems arise
naturally in the description of Josephson junctions arrays,
charge density waves,
multimode lasers, neural dynamics, ecological and evolution models
[Wiesenfeld \& Hadley, 1989; Wiesenfeld {\it et al.}, 1990;
Strogatz {\it et al.}, (1989); Nakagawa \& Kuramoto, 1994; Kaneko, 1994].
Many of these systems can be modeled with
globally coupled maps (GCM)  [Kaneko, 1990]. In particular, GCM's 
can show a variety of collective behaviors such as:
a) formation of clusters, i.e., differentiated subsets
of synchronized elements within the network [Kaneko, 1995];
b) non-statistical fluctuations of the mean field of the 
ensemble [Kaneko, 1995; Perez {\it et al.}, 1993];
and c) global quasiperiodic motion [Kaneko, 1991; Pikovsky \& Kurths, 1994].

Usually, numerical simulations have been performed on the entire
globally coupled system to search for
the occurrence of various types of ordered collective behaviors. 
However, this
direct procedure gives little information
about the mechanism for the emergence of collective behaviors.
In a previous work [Parravano \& Cosenza, 1998] we have proposed 
an alternative method that allows to predict the emergence
of given types of periodic collective behaviors in GCM systems exhibiting
periodic clusters. In this article, we investigate the conditions for 
the emergence of a class of ordered collective behavior in a GCM 
characterized by the occurrence of
$K$ identical clusters, all of them runing periodically among $K$ states but 
out of phase with respect to each other in such a way that the mean 
field of the system remains constant during its evolution.  
This kind of collective behavior has 
been observed for globally coupled logistic maps as 
three identical out of phase clusters by Shinbrot [1994], 
who noticed that the addition of
a constant term to a single local map has a similar dynamics as the GCM system.    
Here we use the similarity between a steadly driven map and
a GCM to predict the conditions for the occurrence of general clustered 
collective behavior in GCM's displaying a constant mean field.

We consider the globally coupled map system
\begin{equation}
\label{ec1}
x_{t+1}^i=(1-\epsilon) f(x_t^i) + \epsilon  \langle x_t \rangle \, ,
\end{equation}
with mean field coupling
\begin{equation}
\langle x_t \rangle = \frac{1}{N}\sum_{i=1}^N x_t^i \, ,
\end{equation}
where $x_t^i$ ($i=1,\ldots,N$)
gives the state of the $i$th element at
discrete time $t$; $N$ is the size of the system; $\epsilon$ measures 
the strength of the coupling; and
$f(x)$ prescribes the local dynamics.
Note that the mean field coupling 
affects all the elements in (Eq.~(\ref{ec1})) in exactly
the same way at all times.
This property of the coupling term allows to establish an analogy 
between the GCM system
[Eq.~(\ref{ec1})] and an associated driven map
\begin{equation}
\label{ec2}
s_{t+1}=(1-\epsilon) f(s_t) + \epsilon L_t\, ,
\end{equation}
where $s_t$ is the state of the driven map at discrete time $t$, $f(s_t)$
is the same local dynamics
as in Eq.~(\ref{ec1}), and $L_t$ is the driving term. In general, $L_t$ 
can be any function of the discrete time $t$. 
Then, any element $x^i_t$ 
in the GCM system behaves equivalently to a single driven map 
if $L_t=\langle x_t \rangle$, $\forall \, t$, and 
their initial conditions are identical,
i.e., $ x^i_o=s_o$. As it will be shown below, the analogy can also
be drawn if $L_t=\langle x_t \rangle$ only for asymptotic times.
The basic idea is that, by studying the asymptotic dynamics of the 
associated driven map, one may infer if specific types of ordered
collective behaviors can emerge in the GCM.  

The simplest situation for which the associated driven map [Eq.~(\ref{ec2})]
can be used to predict the emergence of ordered collective behavior in the 
GCM system [Eq.~(\ref{ec1})] occurs when the mean field 
remains constant, i.e., $\langle x_t \rangle=C$. This behavior takes place in 
the GCM when $K$ identical clusters, each having $N/K$ elements and period $K$, 
are evolving out of phase in order to yield a constant $\langle x_t \rangle$. 
In this case,  the behavior of any of such clusters
in the GCM, can be emulated by an associated driven map subjected
to a constant force $L_t=C$.

The steadly driven map will have a unique asymptotic orbit 
whenever the local map $f$ satisfies Singer's theorem.
This is the case for the logistic map $f(x)=1-ax^2$,
$a \in (0,2]$, which will be used as an example 
to explain our procedure. 
The unique asymptotic orbit of the driven map can be periodic 
on some regions of its space of parameters 
(i.e., in $(a, \epsilon, C)$ for the chosen $f$).
Let us consider an orbit of period $K$, 
and let the sequence of the points on this periodic orbit be 
$\{\sigma_1,\sigma_2,\ldots,\sigma_K\}$. 
By using these $K$ points on the periodic orbit of the driven map, 
the following associated coupling function can be defined
\begin{equation}
\label{eqtheta}
\Theta(a, \epsilon, C) = \frac{1}{K}\sum_{j=1}^K \sigma _j \, .
\end{equation}
The associated function $\Theta$ depends on the values 
$\{\sigma_1,\sigma_2,\ldots,\sigma_K\}$, 
which themselves depend on the value of the constant drive $C$, on the
coupling parameter $\epsilon$, and on the 
local map $f(x)$.
The function $\Theta$ gives the value that the 
mean field in the GCM system 
[Eq.~(\ref{ec1})] would have at a given time
if at this instant its elements 
are segregated in $K$ identical clusters distributed in the states 
$\{\sigma_1, \ldots,\sigma_K\}$.
Note that 
the relation $\Theta=\langle x_t \rangle$ will persist for subsequent times 
if the clusters in the GCM sequentially follow the values 
$\{\sigma_1, \ldots,\sigma_K\}$
while remaining shifted one time step with respect to each other.
The relation $\langle x_t \rangle=\Theta$ is guaranteed 
for all times when the driven map satisfies the condition 
$\Theta=C$. If $C_*$ is the solution of $\Theta(a, \epsilon, C)=C$, 
then the evolution of a set of $K$ independent maps
driven by the same constant force $C_*$
describes a collective behavior in the corresponding
GCM consisting of $K$ identical, period $K$ clusters which maintains 
$\langle x_t \rangle=C_*$.
The driven maps, as well as the clusters in the GCM, will cyclically commute 
over the states $\{\sigma_1(C_*), \ldots,\sigma_K(C_*)\}$. 

It should be noticed that for a given GCM the associated coupling 
function $\Theta$ can be constructed only when the associated driven map 
displays periodic asymptotic responses for some values of $L_t=C$.
As an example, Fig.~1 shows the bifurcation diagram of $s_t$ as function of
$\epsilon$ for the local map 
$f(x)=1-ax^2$ with $a=2$, driven by a constant term $C=0.07$.
Figure~1 also shows  
the time average $\bar{s}$ of the iterates $s_t$ as a function of
$\epsilon$. The statistical quantity $\bar{s}$ gives the value of $\Theta$ 
on the periodic regions of the bifurcation diagram.
The intersection of the curve $\bar{s}$ with the line $C=0.07$ 
in the period three window
gives the value of $\epsilon$ that satisfies the condition
$\Theta (a=2, \epsilon, C=0.07)=0.07$.
Consequently, for this value of $\epsilon$ ($\approx 0.065$) and for $a=2$, 
a GCM system coupled through its mean field can exhibit a 
collective behavior consisting of three identical out of phase clusters 
that yield a constant $\langle x_t \rangle=0.07$.

Note that this prediction is made without direct numerical simulation on the
GCM system. However, the solution of the equation $\Theta=C$
does not tell which initial conditions
on the GCM will conduce towards this particular collective behavior.
We have verified that the GCM system coupled through the mean field reaches
the predicted behavior starting from initial conditions $\{x_o^i\}$
consisting of three groups of $N/3$ elements, each group distributed
around one of the states $\{\sigma_1,\sigma_2,\sigma_3\}$ with some 
dispersion.
For $N=1200$, the system reaches the expected behavior 
when the dispersion is less than $0.007$. 
For larger dispersions or for initially uniform distributed elements on 
the interval $[-1,1]$, 
the GCM system tends to split into three similar but not identical clusters
resulting in a period three mean field that oscillates 
around the value $0.07$ with a small amplitude.
The later case was observed by Shinbrot [1994]   
in a GCM system with a quadratic local map and mean field coupling. 
In general, oscillations of the mean field arise in the clustered 
collective behavior of a GCM when $N$ is not a multiple of $K$.
 
Figure~2 shows $\Theta(a=2, \epsilon, C)$ as 
a function of $C$ for several
values of the coupling strength $\epsilon$.  The continuous curves
correspond to values of $\Theta$ in periodic regions
of the indicated values of $\epsilon$. In the chaotic regions 
$\Theta$ cannot be calculated; however we show $\bar{s}$
for the three smaller values of $\epsilon$ for which chaos can occur. 
The intersections of the curves with the diagonal give the solutions to the
equations $\Theta (a=2, \epsilon, C)=C$ for each
$\epsilon$ shown. 

Figure~3 shows the solution of the equation $\Theta (a=2, \epsilon, C)=C$ 
on the plane of parameters $(\epsilon,C)$ for various regions
where the associated driven map has periodic orbits. The labels on
the different portions of the solution curve indicate the period $K$ of the
orbits. For each value of $\epsilon$ and $C$ on this curve a collective 
behavior 
comprising $K$ periodic clusters that result in a constant mean field 
$\langle x_t \rangle=C$ can take place in the corresponding
GCM for appropriate initial conditions.
The inserts in Fig.~3 are magnifications of the solution $\Theta=C$ 
corresponding to two different periodic windows of the driven map.

In summary, 
based on the analogy between a GCM system and a single driven map,  
we have presented a method to predict the parameter values for the existence of
periodic clustered collective behavior in GCM's displaying a constant mean field. 
By obtaining the asymptotic orbits of the driven map, the associated
coupling function $\Theta$ can be constructed. This function allows to
establish a direct connection between a GCM and its associated driven map. 
The solutions of equation $\Theta(a, \epsilon, C)=C$ represent 
surfaces in the space of parameters $(a, \epsilon, C)$ of the driven map.
For a set of parameter values $(a, \epsilon, C)$ on these surfaces,
the corresponding GCM
can have a mean field $\langle x_t \rangle=C(a, \epsilon)$ resulting
from a periodic clustered behavior. The method also gives the states
$\{\sigma_1(C), \ldots,\sigma_K(C)\}$ sequentially visited by the clusters.
Results were shown for the logistic map with $a=2$, however, the procedure
can be applied for other local maps and for any global coupling function
invariant to permutations of its $N$ arguments.

\section*{Acknowledgment}
This work was supported by the Consejo de Desarrollo
Cient\'{\i}fico, Human\'{\i}stico y Tecnol\'ogico of
the Universidad de
Los Andes, M\'erida, Venezuela.

\section*{References}

\noindent{Chat\'e, H. \& Manneville, P. [1992a]
``Emergence of effective low-dimensional dynamics in the
macroscopic behavior of coupled map lattices",
{\sl Europhys. Lett.} {\bf 17}, 291.} 

\noindent{Chat\'e, H. \& Manneville, P. [1992b]
``Collective behaviors in spatially extended systems with local
interactions and synchronous updating", 
{\sl Prog. Theor. Phys.} {\bf 87}, 1.} 

\noindent{Chat\'e, H., Lemaitre, A., Marq, P.  \& Manneville, P. [1996]
``Nontrivial collective behavior in extensively-chaotic dynamical systems:
an update",
{\sl Physica A} {\bf 224}, 447.} 

\noindent{Chawanya, T.\& Morita, S. [1998]
``On the bifurcation structure of the mean-field fluctuations in the globally
coupled tent map system",  
{\sl Physica D} {\bf 116}, 44. }

\noindent{Kaneko, K. [1990a]
``Globally coupled chaos violates the law of large numbers but not
the central-limit theorem",                                                                                                                                                                            
{\sl Phys. Rev. Lett.} {\bf 65}, 1391.} 

\noindent{Kaneko, K. [1990b]
``Clustering, coding, switching, hierarchical ordering, and 
control in a network of chaotic elements",
{\sl Physica D} {\bf 41}, 137.} 

\noindent{Kaneko, K [1991]
``Globally coupled circle maps", 
{\sl Physica D} {\bf 54}, 5.} 

\noindent{Kaneko, K. [1995] 
``Remarks on the mean field dynamics of networks of chaotic elements",
{\sl Physica D} {\bf 86}, 158.}

\noindent{Nakagawa, N. \& Kuramoto, Y. [1994]
``From collective oscillations to collective chaos in globally coupled
oscillator system",
{\sl Physica D} {\bf 75}, 74.}

\noindent{Parravano, A.  \&  Cosenza, M. G. [1998]
``Driven maps and the emergence of ordered collective behavior in 
globally coupled maps",   
{\sl Phys. Rev. E} {\bf 58}, 1665.}

\noindent{Perez, G., Sinha, S. \& Cerdeira, H. A. [1993]
``Order in the turbulent phase of globally coupled maps",
{\sl Physica D} {\bf 63}, 341.} 

\noindent{Pikovsky, A. \& Kurths, J. [1994]
``Collective behavior in ensembles of globally coupled maps", 
{\sl Physica D} {\bf 76}, 411.} 

\noindent{Shinbrot, T. [1994]
``Synchronization of coupled maps and stable windows",
{\sl Phys. Rev. E} {\bf 50}, 3230. }

\noindent{Strogatz, S. H., Marcus, C. M., Westervelt, R. M. \& Mirollo, 
R. E. [1989]
``Collective dynamics of coupled oscillators with random pinning", 
{\sl Physica D} {\bf 36}, 23.} 

\noindent{Wiesenfeld, K. \& Hadley, P.[1989]
``Attractor crowding in oscillator arrays",
{\sl Phys. Rev. Lett.} {\bf 62}, 1335.} 

\noindent{Wiesenfeld, K., Bracikowski, C., James, G. \& Roy, R. [1990]
``Observation of antiphase states in a multimode laser",
{\sl Phys. Rev. Lett.} {\bf 65}, 1749.} 


\section*{Figure Captions}

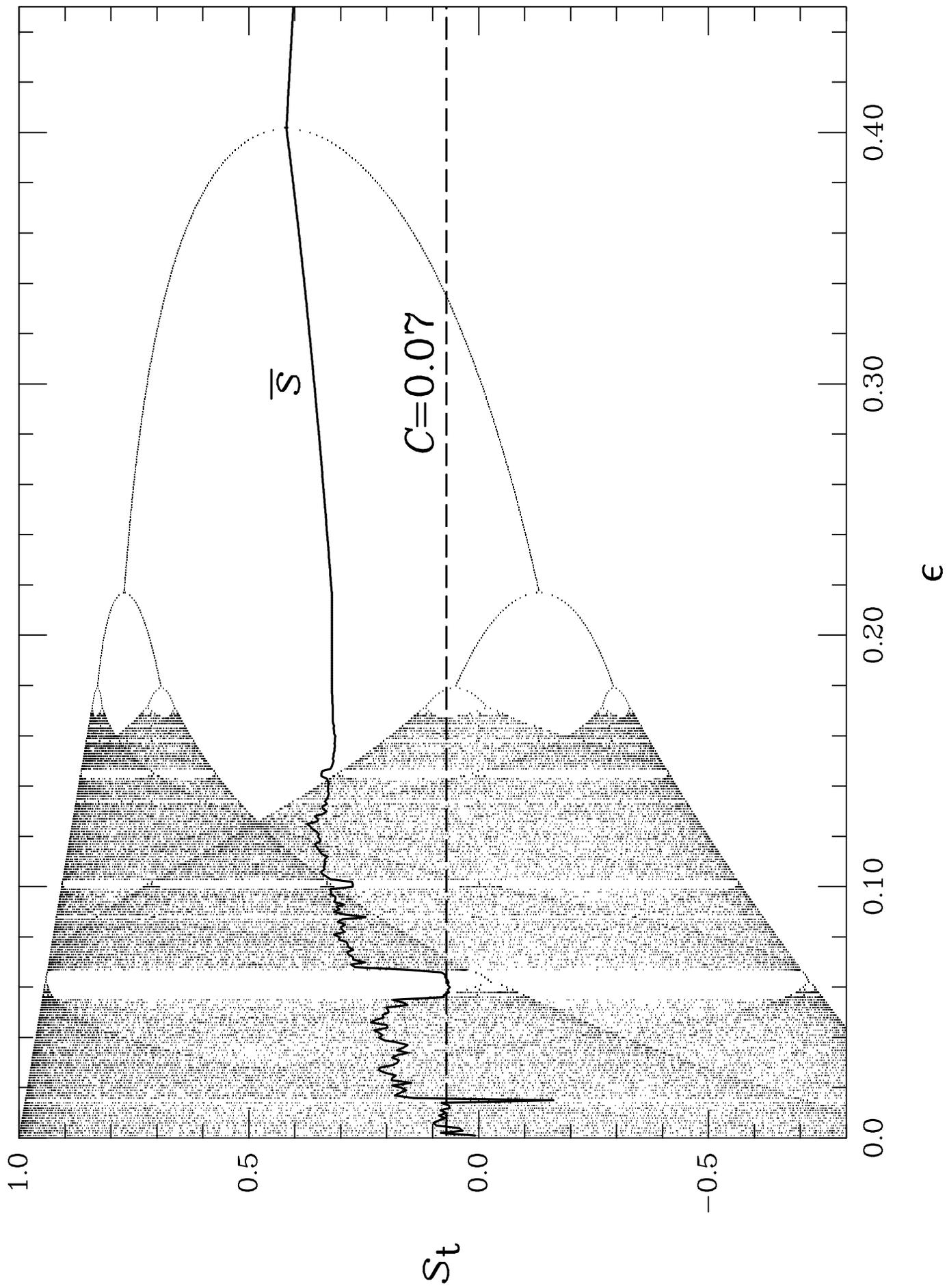
\noindent{{\bf Fig.~1:}} Bifurcation diagram for Eq.~(3) as a function of 
$\epsilon$,
with fixed $a=2$ and $L_t=C=0.07$. The horizontal line corresponds to 
the constant value of $C$. The continuous curve is the time average $\bar{s}$,
which gives the function $\Theta$ on the periodic regions of the diagram.\\

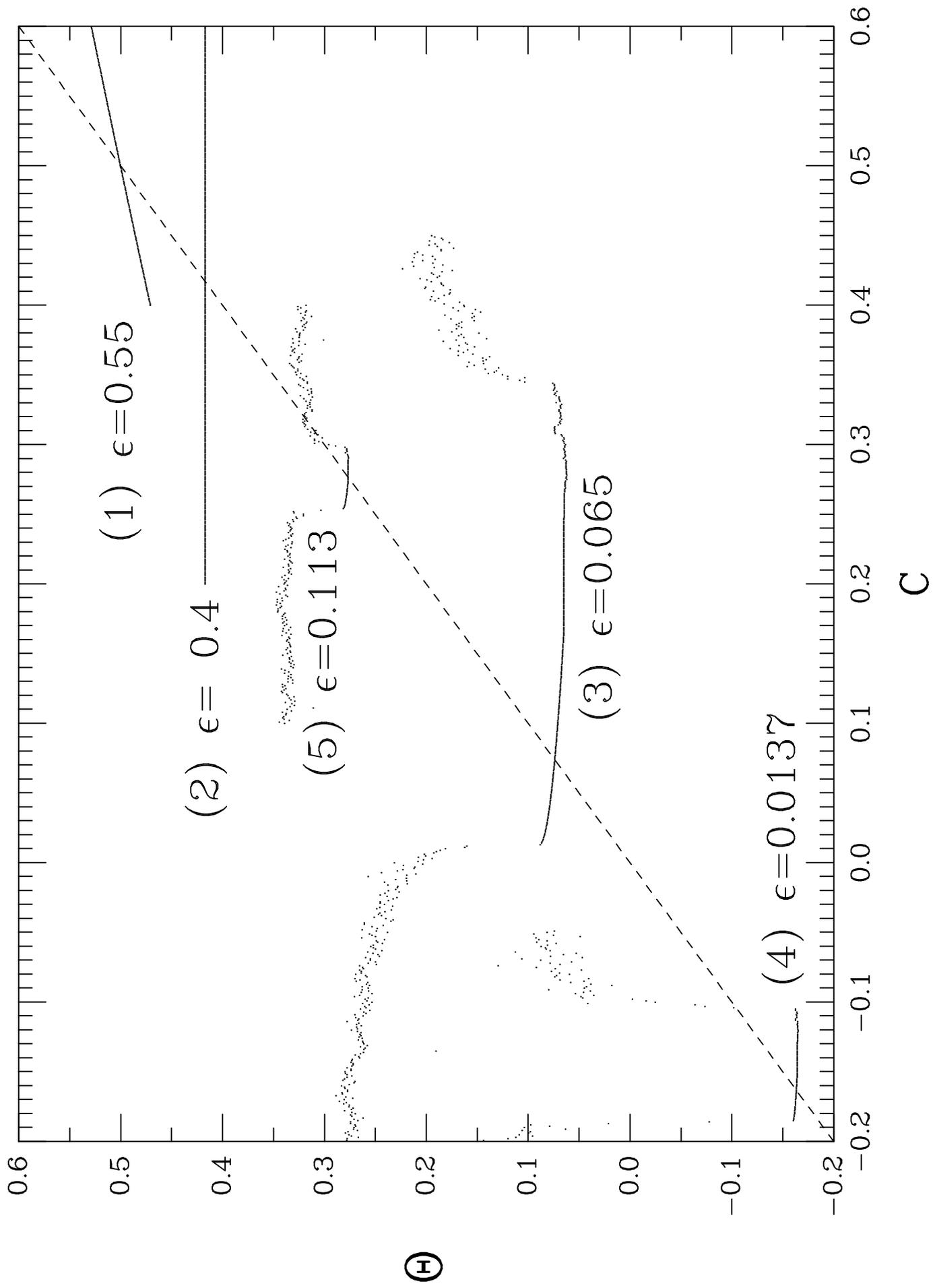
\noindent{{\bf Fig.~2:}} Associated coupling function $\Theta$ 
(continuous curves)
on periodic regions of Eq.~(3), as a function of $C$, for several 
values of $\epsilon$ and fixed $a=2$. The numbers in parenthesis indicate
the periodicity $K$. The dots correspond to values of 
the time average $\bar{s}$ on chaotic regions.\\

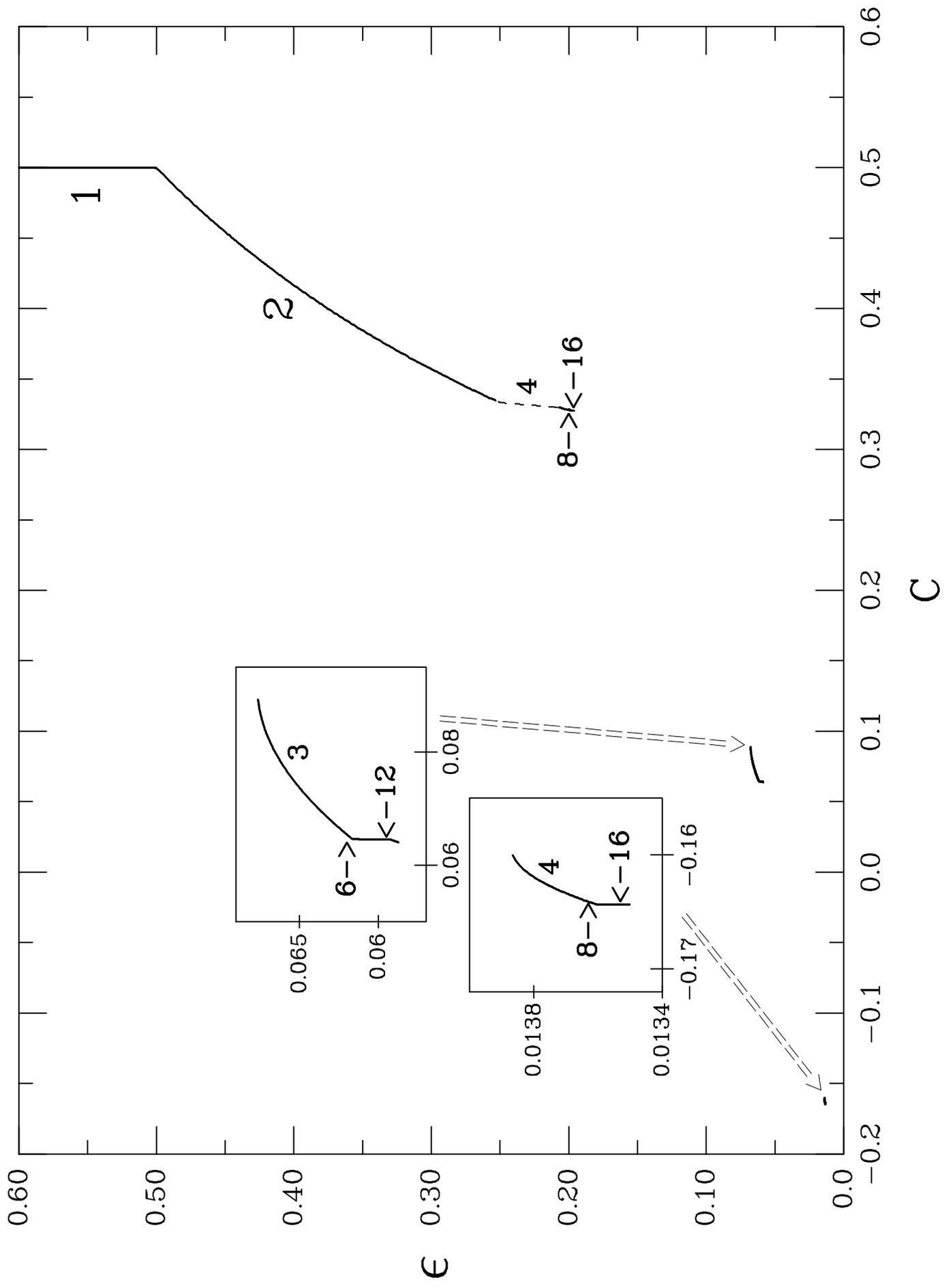
\noindent{{\bf Fig.~3:} The solution
of $\Theta (a=2, \epsilon, C)=C$ for various regions
where Eq.~(3) has periodic orbits. Labels 
indicate the period $K$ of the
orbits.  The inserts are magnifications of the indicated portions
of the curve corresponding to two different periodic windows of 
the driven map}

\end{document}